\documentclass[USenglish,oneside,twocolumn]{article}

\usepackage[utf8]{inputenc}
\usepackage[big]{dgruyter_NEW}
\usepackage{tikz}

\begin{document}

\author*[1]{Rami Ailabouni}

\author[2]{Orr Dunkelman}

\author[3]{Sara Bitan}

\affil[1]{University of Haifa, Israel, E-mail: railaboun@staff.haifa.ac.il}

\affil[2]{University of Haifa, Israel, E-mail: orrd@cs.haifa.ac.il}

\affil[3]{Technion, Israel, E-mail: sarab@cs.technion.ac.il}

  \title{\huge DNS-Morph: UDP-Based Bootstrapping Protocol For Tor}

  \runningtitle{DNS-Morph: UDP-Based Bootstrapping Protocol For Tor}


\begin{abstract}
{Tor is one of the most popular systems for anonymous communication and censorship circumvention on the web, currently used by millions of users every day. This puts Tor as a target for attacks by organizations and governmental bodies whose goal is to hinder users' ability to connect to it. These attacks include deep packet inspection (DPI) to classify Tor traffic as well as legitimate Tor client impersonation (active probing) to expose Tor bridges. As a response to Tor-blocking attempts, the Tor community has developed \textit{Pluggable Transports (PTs)}, tools that transform the appearance of Tor's traffic flow.\\In this paper we introduce a new approach aiming to enhance the PT's resistance against active probing attacks, as well as white-listing censorship by partitioning the handshake of the PT from its encrypted communication. Thus, allowing mixing different PTs, e.g., ScrambleSuit for the handshake and FTE for the traffic itself. We claim that this separation reduces the possibility of marking Tor related communications. To illustrate our claim, we introduce DNS-Morph: a new method of transforming the \textit{handshake} data of a PT by imitating a sequence of DNS queries and responses. Using DNS-Morph, the Tor client acts as a DNS client which sends DNS queries to the Tor bridge, and receives DNS responses from it. We implemented and successfully tested DNS-Morph using one of the PTs (ScrambleSuit), and verified its capabilities.}
\end{abstract}

\maketitle
\section{Introduction}
Internet censorship is the act of inspecting, controlling, and limiting what can be accessed, published, or viewed on the Internet. Organizations and governments engage in Internet censorship due to variety of reasons, among them political, moral, or religious. Since Tor became public in 2002~\cite{FirstTor} and started gaining popularity among users in the world as a system for anonymous communication and censorship circumvention, many countries tried to block their citizens' connections to it. These attempts started with simple methods like blacklisting Tor's website~\cite{TorWebsite} so users would not be able to reach it and download the Tor client software~\cite{TorClientSoftware}, and got more sophisticated with time to include actively downloading the Tor nodes (also called relays) list from the Tor \textit{Directory Servers} and blacklisting them, deploying DPI to search for Tor communication characteristics (e.g., Tor's TLS handshake cipher suite~\cite{GFC2012}), as well as active probing (impersonating a Tor client and connecting to suspicious servers to check whether they run a Tor relay)~\cite{ActiveProbing, KnockingBrigesDoors, GFC2015}.

On the other side of the arms race, the Tor community did not stand still and developed methods to bypass the blocking attempts, mainly the introduction of Tor \textit{Bridges}\footnote{Bridge: a relay which is unlisted in the public directory servers lists. The information needed to connect to a bridge is obtained out-of-Tor (e.g., via BridgeDB~\cite{BridgeDB}, an Email, or in person).} and \textit{Pluggable Transports (PTs)}.

\begin{figure}[h]
    \centering
    \captionsetup{justification=centering, margin=0.5cm}
    \includegraphics[scale=0.525]{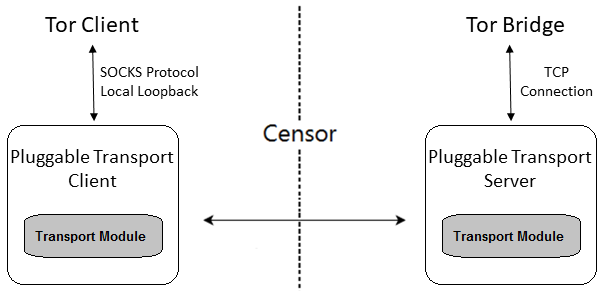}
    \caption{Pluggable Transports Design}
    \label{fig:PT}
\end{figure}

\textit{Pluggable Transports (PTs)}~\cite{PluggableTransportsSpec} are a generic framework for the development and the deployment of censorship circumvention techniques. Their main goal is to obfuscate the connection between a Tor client and a bridge serving as a Tor entry guard, so it looks benign. The PT consists of two parts as seen in Figure~\ref{fig:PT}, one is installed on the Tor client side, and the other is installed on the bridge's side. The PT exposes a SOCKS proxy~\cite{SOCKS} to the Tor client application, and obfuscates or otherwise transforms the traffic, before forwarding it to the bridge. On the bridge's side, the PT Server side exposes a reverse proxy that accepts connections from PT clients and decodes the obfuscation/transformation applied to the traffic, before forwarding it to the actual bridge application.

Data transmitted between a PT client and a PT server can be encrypted, chopped into generic lengths, or otherwise obfuscated in many ways, making it difficult for a censor to detect Tor data and block it. Data transformation/obfuscation and the reverse operations are done by the \textit{Transport Modules/Obfuscation Protocols} used by the PT.

As of Sep. 2017,\footnote{Since Sep. 2017, some changes were introduced to Tor: ScrambleSuit is no longer supported and was replaced by obfs4. Snowflake PT~\cite{Snowflake} is now available in the Tor Browser for some operating systems.} the available and deployed obfuscation protocols in the Tor Browser are: obfs3~\cite{obfs3-spec}, obfs4~\cite{obfs4-spec}, ScrambleSuit~\cite{ScrambleSuit}, FTE~\cite{FTE}, and meek~\cite{Meek}. A brief explanation about each of them can be found in Section~\ref{RelatedWork}. These obfuscation protocols can be divided into two groups: 
\begin{enumerate}
\item Random stream protocols (obfs3, obfs4, and ScrambleSuit). These protocols' communication is shaped as streams of random bytes that cannot be associated with any known protocol. These protocols have two phases: a handshake phase in which the two participating parties securely exchange keys and/or tickets, and a communication phase, consisting of exchange of encrypted messages using the established keys.
\item Structured stream protocols (FTE and meek), which try to mimic known white-listed protocols such as HTTP.
\end{enumerate}

Censoring countries with active probing capability can expose bridges communicating using obfs3. In addition, random stream protocols are of the ``look-like-nothing'' protocols, which means that they can be identified and blocked by a censor using a white-listing strategy, as their fingerprint (including their handshake) does not fit any known protocol (see for example our experiments with DPI tools in Section~\ref{DPITools}).

On the other hand, structured stream protocols that mimic widely used protocols, are resistant to white-listing based blocking. However, they do not protect against active probing~\cite{MeekEvaluation, FteEvaluation}.

Our goal is to improve random stream protocols by using the advantages of structured stream protocols, while maintaining their protection against active probing. We believe that the separation of these protocols to a handshake and communication phase, and encapsulation of their handshake in packets of a known, widely used, white-listed protocol, will strengthen their ability to avoid censorship and detection by DPIs.

DNS was chosen as a protocol to encapsulate the handshake and meet the conditions discussed before for the following reasons:
\begin{enumerate}
\item It is one of the most critical protocols of the Internet~\cite{DNSDomainNames}. Blocking or greatly interfering with this protocol for any reason will induce unacceptable costs on the censoring countries.
\item DNS queries can be automatically relayed from one DNS server to another, until they reach their final destination (recursive DNS queries). This allows relaying data between several DNS servers as an additional layer of protection against connections' tracking and blocking.
\item DNS is a UDP-based protocol. UDP is connection-less and the efforts required to perform DPI of UDP traffic are significantly higher compared to TCP traffic.
\end{enumerate}

We note that the DNS encapsulation is proposed only for the handshake phase.

\subsection{Our Contribution}

DNS-Morph is implemented as an obfuscation layer for random stream protocols. This layer encapsulates the protocol handshake in DNS queries and responses in a way that avoids protocol abnormalities (i.e., a sequence of DNS packets without response, followed by a ``burst'' of responses, a sequence of packets with identical length, oddly structured domain names, etc.). Also, the limited DNS communication avoids tools that defeat IP over DNS tunneling.\footnote{Encapsulating the two phases of a protocol into DNS packets will cause high DNS traffic, which can raise suspicion of DNS tunneling.}

We focus on the handshake phase in this paper because it is the phase DPI tools\footnote{DPI traffic detection is based on pattern/signature matching, anomalies or traffic amount.} and active probes target when searching for Tor traffic and bridges. Also, we later discuss (Section~\ref{FutureWorks}) the possibility of connecting our DNS-Morph to FTE (which shapes Tor's data using regular expressions, and is not active probing resistant) in order to bypass a white-listing censor while resisting active probing.

In order to show the advantages of DNS-Morph, we implemented it in Python and integrated it into Tor's Obfsproxy code~\cite{ObfsProxySourceCodePython}.\footnote{While Tor has moved to PTs in Go~\cite{ObfsProxySourceCodeGo}, we decided to continue using the Python Obfsproxy to easy the development.} As Obfsproxy works only with TCP, and our DNS-Morph works with UDP, we added UDP support to Obfsproxy. We also added acknowledgments and packet reordering on the receiver side, to provide lightweight transport reliability that UDP lacks. This contribution may be of independent interest for PTs that will support UDP communication (e.g., in Marionette~\cite{Marionette}).

We tested our design with the ScrambleSuit protocol in two different environments: a non-censoring environment and a partially censoring environment. Success rates, connection timing, and bandwidth are provided in the experiments Section (Section~\ref{TestsResults}).

Our source code is available online~\cite{DNSMorphSourceCode}.

\section{Related Work}\label{RelatedWork}

\textbf{Protocol Obfuscation}: As of Sep. 2017, five obfuscation protocols were deployed and available to use with Tor as PTs:

\begin{enumerate}
\item Obfs3~\cite{obfs3-spec}: builds an additional layer of encryption over Tor's TLS connection in order to hide its unique characteristics. An un-authenticated customized Diffie-Hellman handshake~\cite{UniformDH} is used to exchange encryption keys. As a result, this protocol is susceptible to active probing attacks.
\item ScrambleSuit~\cite{ScrambleSuit}: protects against active probing attacks by using out-of-band exchanged secrets and session tickets for authentication. ScrambleSuit is also capable of changing its network fingerprint (packet length distribution, inter-arrival times, etc.). This protocol is the predecessor of Obfs4 and is subject to white-listing based censoring.

\item Obfs4~\cite{obfs4-spec}: has the same features as ScrambleSuit, but utilizes the \textit{Elligator} technique~\cite{Elligator} for public key obfuscation, and the \textit{ntor} protocol~\cite{NtorhHandshake} for one-way authentication. This results in a faster protocol than ScrambleSuit and the addition of bridge authentication.
This protocol can also be blocked by white-listing based censoring.
\item Meek~\cite{Meek}: uses a technique called \textit{Domain Fronting}~\cite{DomainFronting} to relay the Tor traffic to a Tor bridge through third-party servers (i.e., CDNs like Amazon CloudFront and Microsoft Azure).
\item Format-Transforming Encryption (FTE)~\cite{FTE}: transforms Tor traffic to arbitrary standard protocols' formats using their language descriptions.
\end{enumerate}
Some other PTs are also available but are not integrated in the Tor Browser. These PTs can be found online~\cite{PluggableTransportsList}.
\\\\
\textbf{Domain Name System}: a hierarchical decentralized naming system that associates various information with domain names on the Internet~\cite{DNSDomainNames}. This information is stored by DNS name servers in DNS records, and can be obtained by sending DNS queries to these servers and receiving their responses.

DNS queries can either be iterative (a DNS resolver client queries a chain of one or more DNS servers, where each server refers the client to the next server in the chain) or recursive (a DNS resolver client queries a single DNS server, which queries other DNS servers on behalf of that client).\footnote{DNS caching is irrelevant to our work as our DNS client sends new DNS queries each time.}

The DNS protocol runs primarily over UDP (TCP is rarely used for client queries).
\\\\
\textbf{DNS Tunneling}: is the act of communicating data of any content inside DNS queries and responses. This technique is used for many purposes, among them bypassing captive portals for paid Wi-Fi services, and command and control or data exfiltration in malware~\cite{DNSMalwareExfiltration, DNSMalwareCandC}.
\\
Three components are used in DNS Tunneling:
\begin{enumerate}
\item Client which sends data in DNS queries and acts like a DNS client.
\item Server which tunnels the client data and sends back DNS responses like a DNS server. This server usually has a registered domain name.
\item Encapsulation mechanism of data into DNS queries and responses, and a corresponding decapsulation mechanism for extracting this data from the DNS queries and responses.
\end{enumerate}

Several DNS Tunneling packages are currently available, some of them are Dnscat~\cite{Dnscat}, Dns2tcp~\cite{Dns2tcp}, and iodine~\cite{iodine}. On the other side, there are techniques for DNS Tunneling detection (e.g., based on the amount of DNS traffic~\cite{DNSExfiltration}).

\section{Threat Model}

Our threat model consists of a \textit{nation-state censor} that desires to block users from connecting to Tor. This censor might use DPI to examine session packets and active probing to check whether suspected servers are Tor bridges or servers. This censor might also be moving towards a white-listing strategy and start blocking access to applications for personal use like Skype~\cite{Skype} or WhatsApp~\cite{WhatsApp}. However, we assume that the censor may not be willing to block fundamental services like HTTP, DNS, IMAP, and FTP, as this can break legitimate communications, thus inducing high economical costs and causing unbearable collateral damage.

We believe this threat model is realistic as recent reports suggest that some censoring countries are continuously tightening their Internet control and taking a comprehensive approach to block all outgoing VPNs traffic during the year of 2018~\cite{VPNBlockChina}.

We also assume that the censor does not perform DNS poisoning. Assumptions about such an attack and possible mitigations are discussed in Section~\ref{FutureWorks} (Future Works).

\section{Obfsproxy Design}\label{ObfsproxyDesign}

Obfsproxy~\cite{ObfsProxySourceCodePython} is an open source software written using Python, and is used by Tor. This software implements the PT design mentioned before, in addition to the obfuscation protocols obfs3 and ScrambleSuit.

Due to the fact that our DNS-Morph is integrated into Obfsproxy, we first describe the Obfsproxy design, and then (in Section~\ref{DNS-MorphDesign}) the modifications done to support DNS-Morph.

\begin{figure*}[h]
    \centering
    \captionsetup{justification=centering, margin=0.5cm}
    \includegraphics[scale=0.5]{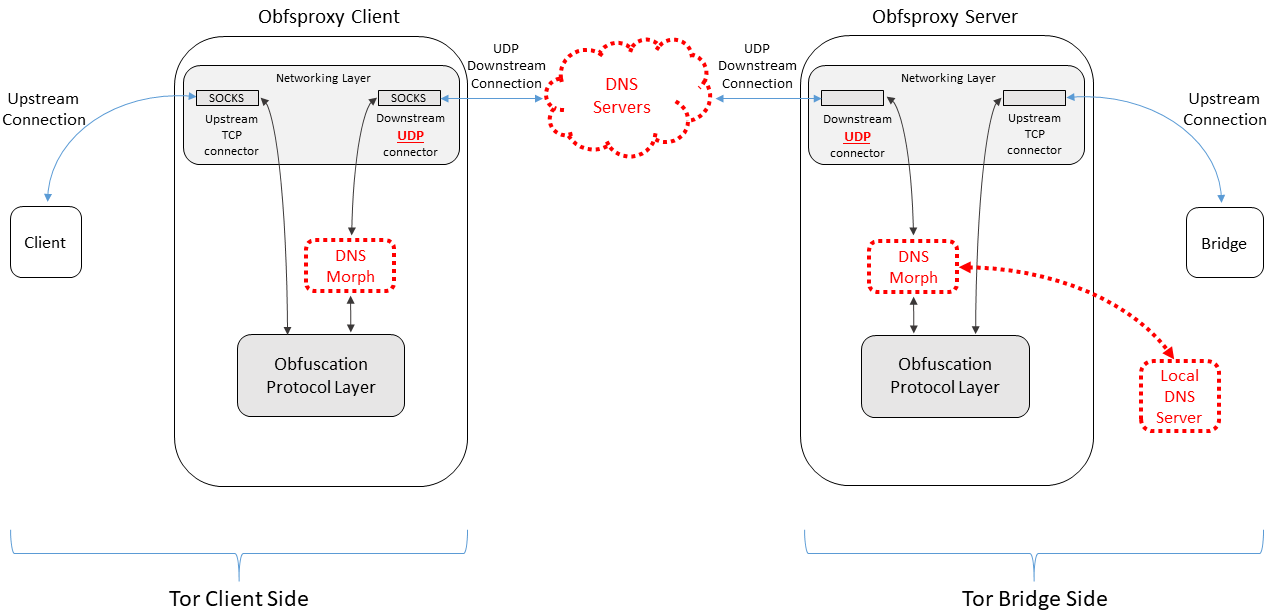}
    \caption{DNS-Morph design: consists of the Obfsproxy design and the new components (dashed red): downstream UDP connector, DNS-Morph, and a local DNS server}
    \label{fig:DNSMorphDesign}
\end{figure*}

Figure~\ref{fig:DNSMorphDesign} (without the red dashed parts) shows the Obfsproxy high level design. Each side, the client and the server, consists of two entities: a Tor client and an Obfsproxy client on the client side, and an Obfsproxy server and a Tor bridge on the bridge side.

Obfsproxy components have two main layers: a networking layer, responsible for connections' establishment, and an obfuscation layer, responsible for the Tor handshake and the data obfuscation.

The connection between the Tor client and the Tor bridge is composed of three components:
\begin{enumerate}
\item  An ``Upstream'' connection between the Tor client and the Obfsproxy client on the client side.
\item A ``Downstream'' connection between the Obfsproxy client and the Obfsproxy server. 
\item An ``Upstream'' connection between the Obfsproxy server and the Tor bridge on the bridge side.
\end{enumerate}

The chronological operation flow of a Tor client and a Tor bridge is as follows:
\begin{enumerate}
\item Client side: launches an Obfsproxy client, which starts a TCP SOCKS listener on its upstream connection.
\\Bridge side: launches Tor bridge and the Obfsproxy server. The Obfsproxy server starts a TCP listener on its downstream connection. The Tor bridge starts a TCP listener on its upstream connection.	
\item Client side: when the Tor client is launched by the user, it connects to the SOCKS TCP listener of the Obfsproxy client on its side. Then, the Obfsproxy client initiates a TCP connection to the Obfsproxy server on the bridge side (downstream connection).
\item Bridge side: when the Obfsproxy server receives the downstream connection request from the Obfsproxy client, it starts its obfuscation layer.
\item Client side: when the downstream TCP connection is established, the client side initiates its obfuscation layer which commences the handshake between the Obfsproxy client and server.
\item If the protocol handshake fails then Obfsproxy modules on both sides close their downstream connection and return to Step 1. If the handshake succeeds, the Obfsproxy server connects to the Tor bridge (upstream connection) and starts receiving data from the Obfsproxy client (downstream connection), decrypting it, and then sending it to the bridge.
\end{enumerate}
At the end of Step 5, the connection between the Tor client and the bridge is fully functional, and the Tor client is connected to the Tor network through the Tor bridge.

\section{DNS-Morph Design}\label{DNS-MorphDesign}

The fact that the handshake phase and the encrypted data exchange phase are two separate phases, this allows us to separate them also is the Obfsproxy design and change properties of each phase without affecting the other one.

We now describe our modifications to the Obfsproxy design to replace its handshake by a DNS-based once.
Our new DNS-Morph design is shown in Figure~\ref{fig:DNSMorphDesign}, and includes three added components (highlighted in dashed red):

\begin{enumerate}
\item Local DNS server: this is a standard DNS server that receives DNS queries and returns DNS responses, used for active probing resistance, as explained in Section~\ref{SecurityAnalysis}.

\item Downstream UDP connector: the connector sends and receives UDP packets. The connector also takes care of the reliability functionality (as discussed in Section~\ref{Reliability}).
 
\item DNS-Morph component:
	\begin{enumerate}
	\item Encodes/decodes data to/from Base32~\cite{Base32RFC}.
	\item Chops data into fragments and encapsulates them into DNS queries or responses.
	\item Decapsulates data fragments from DNS queries or responses, and reassembles these fragments.
	\item Encrypts/decrypts data.
	\item Builds DNS queries and responses.
	\item Communicates with the local DNS server to send and receive queries and responses.
	\end{enumerate}

\end{enumerate}

Our changes to the original Obfsproxy flow of operation in order to support the DNS-Morph flow of operation are:

\begin{enumerate}
\item Downstream connection was changed from TCP to UDP. The Obfsproxy server listens on port 53 UDP, like a standard DNS server\footnote{The Obfsproxy server can be easily configured to listen on other ports, but DNS queries directed to a UDP port other than 53 look suspicious.}~\cite{ServicesPorts}.
\item Obfsproxy client initiates a UDP downstream connection on port 53 to the server. Using the DNS-Morph module, handshake data is encoded to Base32, chopped into fragments of length 20--50 characters,\footnote{~\cite[p. 251]{DNSLength1} suggests that the average DNS query length is 36--59 bytes.~\cite{DNSLength2} suggests that most of the DNS packets that were captured were of size 70--98 bytes, if we subtract from these sizes the size of an empty DNS packet (40 bytes = 20 bytes for IPv4 header~\cite{InternetProtocol} + 8 bytes for UDP header~\cite{UDPHeader} + 12 bytes for DNS header~\cite{DNSDomainNames}) we will get a domain name length of 30--58 bytes.} and encapsulated into DNS queries of type A (address mapping records).

These queries are sent to the Obfsproxy server using direct or indirect routes:
\begin{enumerate}
\item Direct: Obfsproxy client sends the queries directly to the Obfsproxy server's address.
\item Indirect: Obfsproxy client uses the recursive DNS property and sends the queries with a domain name registered by the Obfsproxy server operator. This way the DNS queries travel between different DNS servers until they reach the Obfsproxy server, which adds an additional layer of protection against connections tracking and blocking.

\end{enumerate}
Direct route usage might look anomalous~\cite[p. 28]{fry2009security}, hence, we recommend using the indirect route.
\item The total length in bytes of the handshake data is encoded inside the first DNS packet, so that the receiving side knows when to stop receiving data, and start processing it.
\item The Obfsproxy server buffers each DNS query it receives, until the total length is reached. In order to preserve resemblance to a standard DNS server, the Obfsproxy server sends the received queries to the local DNS server available on its side. When a response is received from the DNS server, the Obfsproxy server sends it back to the Obfsproxy client. When the buffered data forms a complete handshake data, the DNS-Morph module decodes it from Base32, reassembles it, and forwards it to the obfuscation protocol layer, which continues processing it using the specified obfuscation protocol (e.g., ScrambleSuit).
\item After the Obfsproxy client finishes sending the handshake data, it should receive handshake data from the Obfsproxy server. As a result, the Obfsproxy client sends one additional dummy DNS query\footnote{We call a DNS query/response dummy if they contain no protocol handshake data.} so that the Obfsproxy server can start sending back its handshake data as a response to this query. The dummy query is used by the Obfsproxy client to trigger transmission of the handshake data by the Obfsproxy server shaped as DNS responses.
\item When the Obfsproxy server sends back its protocol handshake data, it encodes it to Base32, chops it into fragments of length 20--50 bytes, and encapsulates them into CNAME DNS records (CNAME records can contain textual data and are relatively widely used). Then sends these records encapsulated together with A type records as responses to the dummy type A DNS queries which the Obfsproxy client sent before. Again, the first DNS response packet will include the total length of the Obfsproxy server sent handshake. The Obfsproxy client decapsulates the data from the DNS responses, buffers it, and keeps sending dummy queries until a complete handshake data is received. Then, the Obfsproxy client decodes the received handshake data back from Base32, reassembles it, and forwards it to the obfuscation layer.
\item When the protocol handshake is successfully completed, and tickets and/or session keys are created, both Obfsproxy sides switch from a UDP downstream connection to a TCP connection. The Obfsproxy server side launches a TCP downstream listener on a port exchanged out-of-band and waits for the Obfsproxy client to connect back to it. When the TCP connection happens, the connection between the Tor client and the bridge is fully functional. In case of a protocol handshake failure, both the Obfsproxy client and server close their downstream connection, and the Obfsproxy server continues to behave like a standard DNS server.\\Discussions about the encrypted communication phase after the handshake phase can be found in Section~\ref{FutureWorks}.
\end{enumerate}

\section{DNS-Morph Reliability}\label{Reliability}

DNS-Morph depends on UDP, hence, we cannot rely on the network transport layer to provide reliable data transmission. Therefore, our implementation must guarantee arrival of the sent packets to the receiver's obfuscation layer in the same order that they were sent in, while adding a minimal delay. To achieve this, we use two main methods: the first includes acknowledgments on each received packet by the receiver, and a re-transmission mechanism. The second includes sorting the packets in an appropriate order when they are received so they can be properly processed.

\subsection{Received Packets Acknowledgments}

The receiver sends an acknowledgment each time it receives a packet. If the sender does not get an acknowledgment, it means that the packet or its acknowledgment has failed to reach their destination, so the sender must resend the packet.

We divide the handshake phase into two parts according to the different roles played by the Obfsproxy client:
\begin{enumerate}

\item \textbf{Obfsproxy client sends handshake data:} each time it sends a DNS shaped packet, the Obfsproxy server should send back an acknowledgment in the shape of a valid DNS response to this query.

The Obfsproxy client behaves similarly to the \emph{selective repeat protocol} using a window size of 4 packets, with a few modifications due to the randomized DNS query ID. Obfsproxy client stores the sent packets in a list sorted by their sending order. The query ID serves as a search key to the list. Each time the client receives an acknowledgment it removes the matching packet from the list. If three acknowledgments are received, but none of them matches the first packet in the list, then this packet is resent with a new DNS query ID.

For example: if the Obfsproxy client has sent four DNS packets and the first three acknowledgments it received matched the second, the third, and the fourth packets in the list, then it will resend the first packet. This can be seen in Figure~\ref{fig:DNS-DNSACK}.

\begin{figure}[h]
    \centering
    \captionsetup{justification=centering, margin=0.5cm}
    \includegraphics[scale=0.26]{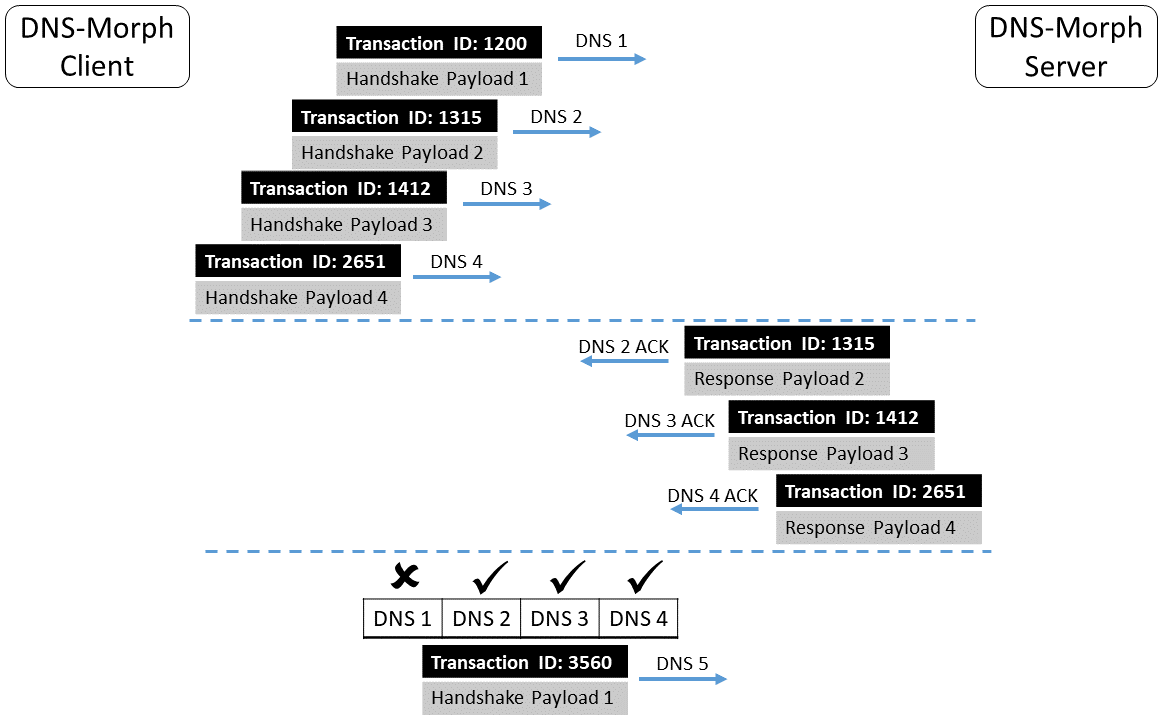}
    \caption{DNS acknowledgments and resending clinet-to-server.}
    
    \label{fig:DNS-DNSACK}
\end{figure}

\item \textbf{Obfsproxy client receives handshake data:} as explained before, each dummy DNS query sent by the Obfsproxy client during this part serves two purposes, the first is to trigger the Obfsproxy server to send one additional handshake packet as a response to the dummy DNS query, and the second is to acknowledge receiving the previous handshake packet sent by the Obfsproxy server.

During this part, the Obfsproxy client implements the \emph{stop and wait protocol}: each time the Obfsproxy client wishes to receive handshake data, it sends a dummy DNS query and waits a while (discussed later) for the Obfsproxy server to send its handshake data as a response to this query. If nothing is received from the Obfsproxy server, then the Obfsproxy client resends the last DNS packet.\footnote{Resending a failed DNS query with the same DNS identifier is a common practice among DNS resolvers.} This procedure is repeated twice (the same packet is sent at most three times in total), and if no responses are received after that, the handshake process is terminated (in a failure).

The time the Obfsproxy client waits for a handshake packet before resending the dummy DNS packet again is calculated by the known weighted average formula for round trip time measuring:

\newtheorem{definition}{Definition}
\begin{definition}{}
Let $RTT_{new}$ be the newly sampled RTT.
\end{definition}
\begin{definition}{}
Let $RTT_{old}$ be the previously calculated RTT.
\end{definition}

Then:
\begin{center}
$RTT = (\alpha \times RTT_{new}) + ((1-\alpha)\times RTT_{old})$
\end{center}
See more details in~\cite[p. 226]{InternetworkingBook}. We use $\alpha=1/8$ as recommended in~\cite{RetransmissionTimer}.

\end{enumerate}

\subsection{Sorting Received Packets}

DNS-Morph adds to each packet a 16-bit identity number which is encrypted (as further explained in Section~\ref{IDEncDec}) and then encoded using Base32 to 4 ASCII characters in the packet's payload.\footnote{The first 4 characters of the payload, as discussed in Section~\ref{DNS-MorphEncodedPackts}.} This number is initialized to the length of the total sent data in the payload of the first DNS packet, and is increased by one each time a packet is sent, so newly sent packets have a bigger identity number than the previously sent ones.

Each time a packet is received, the receiver decodes and then decrypts the DNS-Morph identity bytes of its payload (discussed in Section~\ref{DNS-MorphEncodedPackts}), checks if they form a legal identifying number, and if they do, the packet's data is buffered. When the whole handshake data is received, the receiver sorts the packets according to the identity number and reassembles them back to a real handshake data, passing it to the obfuscation layer for further processing.

In case of receiving two packets with the same identity number, the receiver does not buffer the second received packet but acknowledges the sender about it. This can happen for example because: 1. the first packet's acknowledgment failed to reach the sender, so the sender resends the packet. 2. an attacker is trying to send a packet that contains the same identity number in its payload.

\subsubsection{DNS-Morph Identifiers' Encryption and Decryption}\label{IDEncDec}

Encrypting the DNS-Morph identifier inside the encoded data of the DNS packets relies on an out-of-band shared password (in a similar way as ScrambleSuit and obfs4), which can be exchanged using the out-of-band communication channel of Tor, used to distribute bridge descriptors~\cite{BridgeDB} or any other out-of-band channel the users agree on to exchange bridges descriptors. We note that DNS-Morph can also use the already out-of-band exchanged passwords as long as these passwords are strong.

Our encryption uses 128-bit AES in counter mode. For each identifier, a new key and an IV are created as the output of: HMAC-SHA-256(shared password, X $\vert\vert$ handshake data encoded in the current packet)~\cite{HMACRFC}, when X=0 for Obfsproxy client encryption, X=1 for Obfsproxy server encryption, and $\vert\vert$ means concatenation.

This encryption protects the identifier from being read without the knowledge of the shared key. This protects DNS-Morph from censoring attacks that identify the masked Tor connection by saving all DNS traffic between two connecting entities A and B, extracting the characters that include the identifier, decoding them, and checking whether:
\begin{enumerate}
\item The decoded characters form sequential numbers (identifiers).
\item The decoded characters of the first DNS packet form a number that matches the total length of the data in all the DNS packets.
\end{enumerate}
This fingerprinting attempt exposes DNS-Morph, but as the identifiers are \emph{encrypted}, it fails.

\subsection{DNS-Morph Multiple Sessions Support}

In addition to encrypting the DNS-Morph identifiers, DNS-Morph also supports multiple clients handshake sessions by adding an encrypted 8-bit session ID to the packet's payload. This ID is randomly chosen by the DNS-Morph client and is encrypted together with the DNS-Morph identifiers using the same keys and IVs. The average number of online Tor bridges in the last two years (2016--2017) is about 2600. The average number of Tor bridge users in the same period is about 50,000 users per day~\cite{TorMetrics}. A rough estimation suggests that each bridge serves about 20 users a day. The probability that two of these users connect at the same the time to the same bridge and conduct a DNS-Morph handshake is small. Hence, a 8-bit session identifier seems sufficient to support several handshake attempts simultaneously (although this can be changed if necessary in the future, e.g., letting the bridges to choose this parameter).

\section{DNS-Morph Encoded Packets}\label{DNS-MorphEncodedPackts}

We now show how the encoded handshake data looks inside a DNS query and a DNS response packets.

Consider, for example, handshake data that a ScrambleSuit client sends to the Obfsproxy server. After encoding this data using Base32 we receive encoded data of 450 characters. This data is chopped to fragments of length 20--50 characters, the first fragment of the chopped data looks like this ``ti3zuto4jrz5r22wsu4ar''. To encapsulate this fragment inside a DNS query packet, we need to add encrypted DNS-Morph identity and session identity to it. The key and IV for the encryption of the DNS-Morph identity and session identity are created by:\begin{center} Key $\vert\vert$ IV = HMAC-SHA-256(shared password, 0 $\vert\vert$ ti3zuto4jrz5r22wsu4ar)\end{center} After computing the key and the IV, we use them to encrypt the DNS-Morph identity (which is 450 - the encoded handshake length in characters for the first packet) and the session identity, which is randomly generated by the client (95 for example).\begin{center}``enpin'' = Encode32(AES-CTR$_{K||IV})$(450||95))\end{center} After computing the encrypted DNS-Morph ID and the session ID, they are encoded and then concatenated with ``ti3zuto4jrz5r22wsu4ar'', and with the string ``.bridge.domain'', to create the query:\begin{center}``enpinti3zuto4jrz5r22wsu4ar.bridge.domain''\end{center} which is then encapsulated in a DNS packet and sent to the Tor bridge side. The bridge side, decodes the first 5 characters of the query, decrypts them using the rest part of the query not including ``.bridge.domain'', and the shared password. Then, it finds out that the length of the encoded handshake to receive for session 95 is 450 characters.

The response packet sent by the Obfsproxy server is a DNS response packet as generated by the local DNS server.

\section{DNS-Morph: Security Analysis}\label{SecurityAnalysis}

We now discuss the security of our DNS-Morph design. Since DNS-Morph was built as an additional protection layer for random stream protocols, it inherits their security properties as a carrier protocol for them. The aim of this section is to prove that DNS-Morph does not harm the security properties of the encapsulated protocol but enhances them.

After DNS-Morph finishes the handshake phase the encapsulated protocol will still enjoy the same security properties as before, while its handshake security was enhanced.

We further discuss the security design of DNS-Morph assuming that the encapsulated protocols satisfy some conditions:
\begin{enumerate}

\item They must have a handshake phase.

\item They must be active probing resistant.

\item Their traffic (the handshake and the encrypted data exchange phases) must be indistinguishable from randomness.

\item They must provide data integrity. Providing data authenticity is recommended but is not a must.

\item The traffic exchanged during the handshake should be as short as possible without affecting the security properties of the protocol, as it affects the number of DNS packets exchanged using DNS-Morph (discussed more later).

\end{enumerate}

\subsection{Censor's DPI Capabilities}

We now further elaborate on our assumptions concerning the censor's capabilities.
We assume that the censor has the following DPI capabilities (which are used to detect DNS tunneling) installed in its ISP infrastructure, either on the DNS servers themselves, or on dedicated firewall/routers:
\begin{enumerate}

\item The DPI mechanism can sort DNS packets by their arrival times, build a pseudo-session out of them using a 4-tuple (UDP, IP$_{client}$, IP$_{server}$, Port$_{server}$), and attempt to detect the structure of ``a session''.

\item The DPI mechanism can search the DNS packets payload for any structural data such as counters, flags or words to signal ``start'', ``stop'', ``resend'', etc.

\item The DPI mechanism can consider the length of a DNS query/response and alert if DNS packets lengths regularities/irregularities are detected. For example, alerting about any domain name request longer than 52 characters~\cite{GuyDNS}.

\item The DPI mechanism can consider the number of DNS queries/responses for each 4-tuple or domain name, and alert if this number exceeds a certain threshold, or if the number of queries and responses differs significantly from what can be classified as benign behavior.

\item The DPI mechanism can detect irregular DNS packets sequences, e.g., a series of queries followed by a series of responses.

\item The DPI mechanism can use regular expressions or entropy estimation to detect suspicious DNS packets payloads.

\end{enumerate}

We note that our assumptions grant the censor capabilities that may be stronger than what commercially available DPI systems are performing. We are unaware of any country that combines all these capabilities while performing DPI.

\subsection{DNS-Morph DPI Resistance}

We added to DNS-Morph the following counter-measures that defeat the above DPI threat model, and make it exposure resistant:

\begin{enumerate}

\item DNS-Morph can send its packets in an arbitrary order and sort them on the receiver side, by decrypting the encrypted identifiers, thus, defeating the adversary's ability to detect the order of the packets and the internal session structure.

\item DNS-Morph encodes the encrypted payload inside DNS queries/responses, and concatenates encrypted packet and session IDs. If a DPI searches for any structural data in the payload, it will fail to find any, as all the encoded handshake data looks random.

\item All the DNS queries/responses of DNS-Morph are of random size between 20--50 bytes, which is the size of a standard DNS query/response~\cite{DNSLength1, DNSLength2}, and should not raise any suspicion due to length. Thus, defeating the adversary's attempts to detect length regularities/irregularities.

\item The number of DNS-Morph exchanged queries/responses depends on the encapsulated protocol. If we take ScrambleSuit for example, we can make some modifications (i.e., reduce its padding size, discussed in Section~\ref{TestsResults}), which will not affect the protocol security traits when DNS-Morph is added, but can reduce the total amount of DNS exchanged packets to as little as 26 DNS packets, an amount not so far from the number of DNS packets exchanged when visiting popular websites such as Google (10--17 packets), YouTube (11--13 packets), Facebook (24--27 packets), Yahoo (60 packets), and Reddit (40 packets), which are among Alexa's top 10 global websites for the year 2017~\cite{AlexaTop}.
Registering more than one domain name on the DNS-Morph server IP (or two IP addresses controlled by the bridge) is an additional mitigation that minimizes the number of DNS packets sent to the same domain name. As discussed in Section~\ref{TestSetup}, this approach is cheap and is also relatively easy to configure.

\item The Obfsproxy server cannot send a DNS response without receiving a DNS query. A small delay can be added while the Obfsproxy client sends the DNS queries, which can eliminate DNS sequence abnormalities.

\item It is hard to use regular expressions in order to check for the ``validity'' of domain names without getting a lot of false positives, as many domain names nowadays can look random (or have a random looking part). Examples can be seen while browsing to ``https://edition.cnn.com/'', a DNS type A query having the content ``d3c8wvwvehjsxe.cloudfront.net'' is sent by the DNS client. An additional example can be seen while browsing to ``https://www.yahoo.com'', a type A DNS query including the content ``shim.btrll.com'' is sent to the DNS server, and a DNS response including two records is received, one of these records include the CNAME ``d3qdfnco3bamip.cloudfront.net'' can be seen. Responses like that can be received as a result of load balancing services deployed to distribute the traffic to the most available server to handle the request. Even if we assume that a censor uses a regular expression to check domain names ``validity'', one can take this \textbf{same} regular expression and use it to encode DNS-Morph data, e.g., using FTE~\cite{FTE}.
We compare CloudFront domain names with DNS-Morph produced ones in Section~\ref{Entropy}.
\end{enumerate}

\subsection{Additional Attacks and Resistance}

In addition to using DPI, a censor can tamper with DNS packets payload in many ways, such as:
\begin{enumerate}

\item The censor can change uppercase letters to lowercase, and vice versa. This change should not affect real DNS as it is case-insensitive, but might affect encoded and encapsulated data as changing cases changes bytes of the real data. As discussed in Section~\ref{DNS-MorphDesignConsiderations}, our method of encoding and decoding data is resilient to this kind of changes.

\item The censor can change the DNS packet payload data or some of it (e.g., DNS poisoning). While deeming DNS poisoning out of this research scope, it is important to note that DNS poisoning in this context is a denial of service attack, i.e., it will cause a handshake failure, but will not reveal more information about the exchanged data than the information that was revealed by the original handshake. This means that if, for example the data exchanged is encrypted using an out-of-band key, DNS poisoning will not disclose the fact that a Tor communication happened, nor the keys used to encrypt it.

\item The censor can inject DNS packets in two different ways:
\begin{enumerate}

\item The injected DNS packet contains random data. The receiver will not consider this packet as a part of the handshake as the possibility of this data to include an encrypted identifier using the out-of-band keys, is small.

\item The injected DNS packet is a replayed packet. As explained in Section~\ref{Reliability} this packet will be considered as a network reliability issue and will not be considered as part of the exchanged handshake.
\end{enumerate}

\item The censor can observe the DNS responses sent back by the server, and try to launch an HTTP session to the IP written in these responses in order to examine if this IP points to a real server. Assuming that the IP returned by the DNS server is the IP of the DNS-Morph server, this server can run a simple HTTP server that displays (or otherwise redirects to) a random web-page, for example, an HTML page is returned each time including a random ``fortune cookie'' wisdom saying/quote (i.e., an HTTP fortune daemon).

\end{enumerate}

\subsection{Active Probing and Replay Attack Resistance}

Our new Obfsproxy server acts like a real DNS server. Each time a DNS query is received, a real DNS response is sent back by a real DNS server. Assuming that the obfuscated protocol is active probing resistant, the probe will send DNS packets and receive real DNS responses. If the probe does not know the out-of-band shared password/key between the real Tor client and bridge, the obfuscated protocol on the Obfsproxy server side will reject the probe's handshake request, while the Obfsproxy server will continue to respond as a real DNS server.

The same also holds with respect to replay attacks assuming that the obfuscated protocol is replay-attack resistant. The obfuscated protocol will reject any replayed packet, while the Obfsproxy server continues to respond to it as a real DNS server.

\subsection{Domain Names' Entropy}\label{Entropy}

To show that the domain names used by DNS-Morph have the same entropy capacity as of regular domain names used in CDNs we have performed a simple test. We took 32 DNS packets directed at CloudFront services and 36 DNS packets obtained from a DNS-Morph handshake. We applied gzip and bzip2 to a file containing only the prefix of the domain names. The compression ratio using gzip of the CloudFront domain names was 66.5\% whereas of the DNS-Morph domain names was 67.3\%. In the case of bzip2, the ratios are 70\% and 71.2\%, respectively. 

Given the fact that the handshake is done once per session, it is easy to see that attacks based on estimating the entropy in the domain names are unlikely to offer high precision due to the small difference in compression rates (suggesting somewhat related entropy capacity)~\cite{Paxson}.

\section{DNS-Morph Design Considerations}\label{DNS-MorphDesignConsiderations}

\subsection*{Choice of DNS}

We now discuss various trade-offs we made in DNS-Morph's design, and explain our design decisions: First, as mentioned before DNS is an essential Internet protocol, and its complete blocking is highly unlikely due to the high cost of doing so, namely, practically disconnecting from the internet. In addition, DNS supports a variety of query types which gives us more freedom in choosing the record type that can encapsulate our encoded data.

Following the selection of DNS, we had to use UDP (TCP sessions in DNS are usually used for zone transfers between DNS servers). This resulted in the need to add a reliability layer for DNS-Morph as loss of even a single packet of handshake data on any side causes a handshake failure (discussed more in Section~\ref{Reliability}).

In addition, the following actions were taken to enable DNS-Morph to mimic (as much as possible) an ordinary DNS communication:

We decided to add dummy DNS queries/responses which do not carry handshake data on both sides during the handshake phase, in order to prevent DNS protocol anomalies where the client sends multiple queries, and the server responds with multiple answers after all the requests were received, or the Obfsproxy server sending DNS responses without any sent queries.

\subsection*{Choice of Base32}

We chose Base32 for encoding the handshake data which can include any byte value into characters that follow the domain name system rules~\cite{DNSDomainNames}:
\begin{enumerate}
\item Domain name can include labels and the character ``.''.

\item Labels must start and end with a letter or a digit, and have as interior characters only letters, digits, and hyphens.

\item Letters can be any one of the 52 alphabetic characters ``A--Z'' in uppercase and ``a--z'' in lowercase.

\item Digits can be any one of the ten digits ``0--9''.

\end{enumerate}

While we could have used Base64~\cite{Base32RFC} or Base58~\cite{Base58}, DNS is not case sensitive. In such encodings, a censor can rewrite every single DNS query to a lowercase one, which does not harm normal DNS requests, but breaks DNS-Morph.
These factors narrow our encoding options to a base that has uppercase letters or lowercase letters, but not both. Base32 includes the alphabetic characters ``A--Z'', the digits ``2--7'' and the character ``='', which is suitable for our uses after changing it to the digit ``1''. We also changed the uppercase letters that Base32 produces to lowercase letters inside the DNS queries/responses in order to make them look more consistent with regular DNS queries/responses. All the letters on the receiver side are converted back to uppercase before the Base32 decoding process, which makes our design resistant to attacks like the one described before.

We note that we can also use Base36, which includes the alphabetic characters ``A--Z'' and the digits ``0--9''. Using Base36 can protect DNS-Morph against a censor with the ability to spot the lack of the digits ``0, 8, 9'' from DNS-Morph packets, as these digits can appear in DNS labels, but are not included in our modified Base32. However, we decided to not use this for the ease of implementation. Future works may wish to explore that approach.

\subsection*{Query Types}

In our choice of which DNS query/responses types to use for handshake data encapsulation, we used the following guidelines:
\begin{enumerate}
\item The type of the DNS query/response must accept by their definition the amount and the character set of the data we want to encapsulate inside them. For example, the type A query can include characters and numbers (domain names), but the type A response includes 32-bit IPv4 addresses, thus, encapsulating encoded data that does not look like a valid IPv4 address inside a type A response is not possible.
\item Using certain DNS query/response packet types must not contradict the general DNS query/response types statistics over the Internet.
\item The types of the queries and responses must match.
\end{enumerate}

To send the Obfsproxy client handshake data, we use type A queries, and receive type A responses from the Obfsproxy server.
To send the Obfsproxy server handshake data, we use CNAME and A DNS records, which are encapsulated together in the same DNS packet, and sent as DNS responses for type A DNS queries.
We chose these types because they meet the aforementioned conditions, and in the same time are widely used on the Internet. The encapsulation of these records in DNS packets is done using the ``dnslib'' Python library~\cite{dnslib}, aiming to build DNS packets that look as real DNS client/server packets.

\subsection*{Recursive DNS}

Finally, we decided to use the recursive DNS property and send the queries with a domain name registered by the Obfsproxy server operator, in order to protect against packets tracking. This also enables to bypass networks' firewall rules which forbid a direct DNS packet to be sent outside the network, besides the network's own DNS server.

\section{Tests And Results}\label{TestsResults}
\begin{figure*}[h]
    \centering
    \captionsetup{justification=centering, margin=0.5cm}
    \includegraphics[scale=0.5]{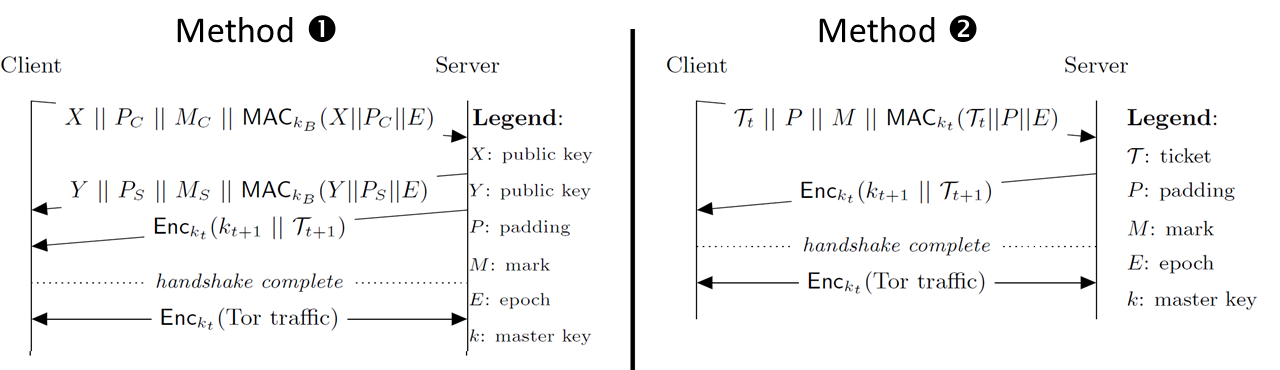}
    \caption{ScrambleSuit's handshake methods.}
    \label{fig:ScrambleSuitHandshake}
\end{figure*}

We have implemented our DNS-Morph design using Python, and tested it successfully with the ScrambleSuit PT. ScrambleSuit has two handshake methods, depicted in Figure~\ref{fig:ScrambleSuitHandshake}.

\newcommand*\circled[1]{\tikz[baseline=(char.base)]{
            \node[shape=circle,fill,inner sep=1pt] (char) {\textcolor{white}{#1}};}}   
            
Using Method~\circled{1} a client and a server generate 4096-bit even private keys $x$ and $y$, respectively, and the corresponding public keys $X=g^x(mod~p)$, $Y=g^y(mod~p)$. After exchanging the public keys, they agree on a shared private key $k_{t}=Y^x(mod~p)=X^y(mod~p)$. Then, using the shared key, the server sends back an encrypted ticket and a new private key, which the client will use for their future connections with handshake method~\circled{2}. An out-of-band shared key $k_B$ is used to calculate the MACs and the marks (used to facilitate the localization of the MACs).

Our tests were performed on ScrambleSuit using method~\circled{1}, as we wanted to simulate a newly connected client, furthermore, a larger amount of data is sent when using this method, thus, making it harder to defend.

\subsection{Test Setup}\label{TestSetup}

For our experiments, we set up Tor clients in two different environments, more information about these environments can be found in Section~\ref{Environments}. These clients connect to a Tor bridge residing at a university located in Israel which has two types of ISPs, ones that censor the Internet and the others do not. The bridge is connected to the Internet using an ISP \textbf{not known} for any censoring activities.
The two sides (the client's sides and the bridges' sides) had the Obfsproxy software with the DNS-Morph component, and tried to establish Tor's first connection to a bridge. We ran 30 connection attempts in each test, measuring rates of successful handshakes, their timing, and their bandwidth.

Between connection attempts, all the temporary files and session states generated by previous attempts (e.g., session tickets) were deleted to ensure that ScrambleSuit generates everything and behaves the same all the time.\\With this setup, we performed three different experiments:
\begin{enumerate}
\item ``Original Design'' tests, to test the original Obfsproxy without DNS-Morph. This Obfsproxy uses TCP only.

\item ``DNS-Morph Direct'' tests, to test the Obfsproxy including DNS-Morph. The Obfsproxy client connections to the Obfsproxy server were direct, i.e., DNS queries were sent from the client directly to the Obfsproxy server address.

\item ``DNS-Morph Indirect'' tests, to test the Obfsproxy, including DNS-Morph. The Obfsproxy client connections to the Obfsproxy server were indirect, and DNS queries were sent indirectly to the Obfsproxy server address by other DNS servers (using recursive DNS queries).

\end{enumerate}

For our ``DNS-Morph'' tests, we also did one additional type of tests: we decreased the maximal size of the random padding added to the protocol (see Figure~\ref{fig:ScrambleSuitHandshake}). ScrambleSuit picks a random padding length of 0--1308 bytes to change the protocol's signature. Due to the use of DNS queries, such a protection is less needed. Thus, we tested the system when the maximal size of the random padding is at most 100 bytes (compared with 1308 originally).

Decreasing the padding length range reduces the \textbf{maximal} total number of DNS packets (queries and responses) per DNS-Morph handshake from 458 packets to 96 packets and the \textbf{average} total number from 262 packets to 81 packets (the \textbf{minimal} total number stays 26 packets). By doing so, we wanted to check if the handshake time improves, and if as a result of DNS servers offloading (DNS-Morph indirect), the handshake success rates will be higher compared to the first type tests, as DNS servers will be more willing to serve us.

It is also worth mentioning that technically, the only difference between running our Obfsproxy client and server versions for the original designs tests and the DNS-Morph direct tests is the flags they receive from the command line when they are launched.

The difference between running the DNS-Morph direct and indirect tests is the domain name we bought for our Obfsproxy server (for an annual price of 0.88 USD~\cite{Namecheap}), configuring the name server of our domain name to point to our Obfsproxy server's IP, and changing one flag from ``direct'' to ``indirect'' when launching the Obfsproxy client and server.

More technical information about running and testing our DNS-Morph version of Obfsproxy can be found in the research Github repository~\cite{DNSMorphSourceCode}.

\subsection{Clients' Testing Environments}\label{Environments}

We setup two Tor clients in two different environments to test DNS-Morph behavior in a censoring environment and in an environment not known for censorship.

\textbf{Environment 1:} This environment is a city in Italy. Italy is \textbf{not known} for acts of Tor censoring. In this environment our Tor client had no problems in connecting to directory servers, obtaining relays descriptors, and connecting to these relays. The Tor client also succeeded while connecting to a Tor bridge using any of the obfuscation protocols.
By setting up a Tor client in this environment, we wanted to test if DNS-Morph works well (including its reliability functionalities discussed before) in a normative environment which does not seek to block it in any way.

\textbf{Results for Environment 1 (non-censoring)}: We performed numerous connection tests during a period of 3 months (May--July 2017). Table~\ref{tab:Times1} contains the results of the tests done on the 5th of July 2017, as a typical example for these results. We chose to present the results of only one day as all the other days produced similar results.

Taking into consideration the fact that no censoring acts were observed, and the relatively large distance between Israel and Italy (about 2500~km), the success rates when DNS-Morph data was relayed between DNS servers (DNS-Morph indirect) are lower (90\% and 93\%) than the success rates when DNS-Morph data was sent directly to the Tor bridge (100\%). Also, we can see that when the DNS-Morph indirect relayed more data, the success rates (90\%) were lower that the success rates when the DNS-Morph relayed the shortened version of ScrambleSuit 93\%) as discussed before.

\begin{table*}[h]
\begin{tabular}{l|cccc|c|c} 
 \hline
 &\multicolumn{4}{c|}{Time {\small(seconds)}}&\\
 &Minimum&Maximum&Average&Median&Success Rate&Bandwidth (bytes)\\
 \hline
{\small Original ScrambleSuit}&0.532&0.735&0.601&0.589&100\%&3705\\ 
{\small Original ScrambleSuit with DNS-Morph Direct}&1.206&5.378&2.952&2.733&100\%&22054\\
{\small Original ScrambleSuit with DNS-Morph Indirect}&1.825&15.727&6.132&5.463&90\%&22840\\
{\small Shortened ScrambleSuit with DNS-Morph Direct}&1.077&2.788&1.968&1.979&100\%&6070\\
{\small Shortened ScrambleSuit with DNS-Morph Indirect}&1.643&5.441&2.971&2.539&93\%&6743\\
 
 \hline
\end{tabular}
\\\\
\begin{scriptsize}
{ScrambleSuit - Original: ScrambleSuit with the original padding (0--1308 bytes).\newline}
{ScrambleSuit - Shortened: ScrambleSuit with a shortened padding (0--100 bytes).}
\end{scriptsize}
\captionsetup{justification=centering}
\caption{Times and success rates of handshakes in Environment 1 (non-censoring).}
\label{tab:Times1}
\end{table*}

\textbf{Environment 2:} This environment includes an ISP in Israel, which is \textbf{well known} for censoring Internet browsing for its customers (who seek this type of censorship). Content censored by this ISP can be pornography, violence, live casting, and videos. While trying to connect to websites like Google, YouTube, and Facebook, this ISP asked us to install CA certificates on the client's machine. Refusing to do so resulted in blocking client's access to these websites. Other websites were simply blocked without even giving the client an option to install certificates (i.e., Yahoo, Bing, and The Tor Project website).

It is worth mentioning that the certificates needed to be installed by this ISP are self-signed certificates issued to ``Netspark''~\cite{Netspark}, a company providing real-time browsing data inspection and web content filtering services.

Connecting to Tor in this partially blocking ISP: when trying to connect in the default method, we encountered failures during different phases such as connecting to the directory servers, loading relay descriptors, or connecting to the entry relay.
Trying to connect using obfs3 or FTE always failed.

Choosing obfs4 succeeded sometimes and on the other times it failed. Even on successful attempts a lot of the bridges were blocked.
Connecting to another ISP from the same test machine gave us a behavior similar to ``Environment 1''.

In conclusion, we suspect that this ISP continuously collects information about Tor bridges and blocks them.  

\textbf{Results for Environment 2 (censoring)}: We performed numerous connection tests during a period of 2 months (January--February 2018). Table~\ref{tab:Times2} summarizes the results of the tests done on the 10th of February 2018, as a typical example for these results. We chose to present the results of only one day as all the other days produced similar results.

Despite the fact that the Tor client was connecting to the Tor bridge using a censoring ISP, this ISP could not block its connection, not of the original ScrambleSuit, nor of our DNS-Morph version. This indicates that DNS-Morph did not harm the original security properties of ScrambleSuit.

The 100\% success rates in all the experiments can be explained by the fact that the client and the server are located in the same country but connected to different ISPs. The distance between the Tor client and the Tor bridge in this environment is about 50~km (in comparison to about 2500~km in Environment 1). The distances in both environments also affect the timing results between them, as we can see that the timing of the same experiment is less in Environment 2 than it in Environment 1.

Further testing in more censoring countries is essential, as discussed in Section~\ref{FutureWorks}. 

\begin{table*}[h]
\begin{tabular}{l|cccc|c|c} 
 \hline
 &\multicolumn{4}{c|}{Time {\small(seconds)}}&\\
 &Minimum&Maximum&Average&Median&Success Rate&Bandwidth (bytes)\\
 \hline
{\small Original ScrambleSuit}&0.044&0.101&0.087&0.091&100\%&4042\\ 
{\small Original ScrambleSuit with DNS-Morph Direct}&0.891&1.813&1.113&0.953&100\%&22140\\
{\small Original ScrambleSuit with DNS-Morph Indirect}&1.844&2.391&2.171&2.189&100\%&24350\\
{\small Shortened ScrambleSuit with DNS-Morph Direct}&0.438&0.672&0.542&0.547&100\%&6528\\
{\small Shortened ScrambleSuit with DNS-Morph Indirect}&0.594&0.798&0.708&0.719&100\%&6621\\
 
 \hline
\end{tabular}
\\\\
\begin{scriptsize}
{ScrambleSuit - Original: ScrambleSuit with the original padding (0--1308 bytes).\newline}
{ScrambleSuit - Shortened: ScrambleSuit with a shortened padding (0--100 bytes).}
\end{scriptsize}
\captionsetup{justification=centering}
\caption{Times and success rates of handshakes in Environment 2 (censoring).}
\label{tab:Times2}
\end{table*}

\subsection{Deep Packet Inspection Tools}\label{DPITools}

To evaluate DNS-Morph resistance against DPI we chose two open source tools: nDPI~\cite{nDPI} version 2.2.0, and Libprotoident~\cite{Libprotoident} version 2.0.12. We captured the packets transmitted between the Tor client and the Tor bridge from the previous tests (Section~\ref{TestsResults}) using Wireshark and ran the two tools to analyze them.

\textbf{Results:} The original Scramblesuit connections were analyzed as ``Unknown TCP'' protocol packets by both tools, nDPI and Libprotoident. The DNS-Morph connections were analyzed by nDPI as ordinary DNS packets for the handshake phase, and SSL packets for the encrypted data exchange after the handshake was done. Libprotoident on the other hand analyzed the DNS-Morph connections as ordinary DNS packets for the handshake phase, and ``Unknown TCP'' protocol packets for the encrypted data exchange after the handshake.

\section{Summary}

In this work we described DNS-Morph, a method to hide PT's handshake communication in a series of DNS queries and replies. We implemented the system and checked that it successfully establishes a Tor connection between a Tor client and a Tor bridge.

The use of DNS offers several layers of security for this process: DNS blocking (or even strong manipulation) comes at a huge price for the censor, DPI attacks on DNS are harder to implement, and DNS enjoys an inherent resilience to blocking attempts due to the nature of recursive DNS queries.

In addition, we have added counter-measures designed to defeat UDP sessions' tracking mechanisms which, we believe still do not exist today in commercial DPI products.

While we have tested the implementation using the ScrambleSuit handshake, it is easy to see that this methodology could work with any PT that satisfies the conditions stated in Section~\ref{SecurityAnalysis}. As already stated, after the handshake, one can transform the remainder of the Tor communication to a white-listed protocol using tools like FTE or meek to maintain white-listed behavior (while enjoying the security against active probing offered by the protected handshake).

\subsection{Future Works}\label{FutureWorks}

This research can be further developed to include more interesting topics and answer currently open questions:

\begin{itemize}

\item We are aware of the DNS poisoning some countries conduct not only to block sensitive content but also to promote local websites~\cite{GFCDNS}. An interesting approach is to examine what exactly is done by the censor while performing DNS poisoning, for example, does the censor only send poisoned responses to DNS clients? Does it also block the DNS query from reaching the real DNS server? Does it permit it and block its DNS response? Future work will test whether our DNS-Morph can finish the handshake despite these DNS poisoning attacks. As noted before, these attacks will not reveal Tor communication.

\item Our security analysis was based on assumptions we collected from the literature about DPIs and DNS tunneling detectors, in addition to examining DNS-Morph against open source DPI products such as nDPI~\cite{nDPI}, and Libprotoident~\cite{Libprotoident}. Obviously, it would be better to test DNS-Morph in real life censoring countries (as current tests were done in somewhat controlled environments).

\item In order to better mimic the behavior of a DNS client (and fulfill one more of the requirements for building a successful mimicking protocol~\cite{Parrot}), a service can be installed on the client machine to collect data about the user's browsing behavior, such as: DNS packets timing, size, DNS client behavior if its query was not answered, etc.
When the DNS-Morph client is started, it can analyze the collected data to better adapt the specific user and machine behavior.

\item After DNS-Morph finishes its handshake, the Tor data exchange phase starts, in which a TCP port is opened in the Obfsproxy server side and encrypted Tor data is exchanged. This could also look suspicious for a DPI. One possible solution is to fake a TLS handshake between the Obfsproxy client and server after a TCP port is opened and before the exchange of the Tor encrypted data. This way the communication looks as ordinary DNS communication followed by an HTTPS one.

\item Currently, DNS-Morph works with A and CNAME DNS record types. Adding more DNS types will make it harder for a DNS tunneling detector to mark our DNS traffic as DNS tunneling traffic. Also, dummy queries could be made less randomly built to raise even less suspicion.

\item Currently, FTE does not use a pre-shared key to encrypt Tor data. The key used in FTE is either:\begin{enumerate}
\item Hard-coded in FTE code.
\item Negotiated using Diffie-Hellman.
\end{enumerate}
Both options are not active probing resistant. One way to overcome this issue can be to initiate DNS-Morph before FTE, with a handshake like the one of obfs4 or ScrambleSuit. After the handshake is successfully completed, Tor obfuscated data can be tunneled with FTE. This communication will look like a DNS interaction at first, and then like an HTTP one, and is expected to succeed in bypassing a fully white-listing censor, and also be active probing resistant. Future work will examine this issue.

\item At the moment DNS-Morph is using Base32 encoding (with small modifications) for the domain name. This may suggest somewhat irregular domain names, which may cause suspicion. One possible future work direction is to use an encoding scheme which mimics as best as possible real domain names. This line of research requires first constructing the regular expressions associated with such names, and then simply using it for embedding the domain names.

\item One possible extension to DNS-Morph is to use multiple DNS servers which collaborate. Namely, a future development route for DNS-Morph is to run 2 (or more) DNS servers, each accepting some of the queries of the handshake (and sending back the information). This way, the tracing of the DNS queries becomes harder (as the number of queries per DNS server is
reduced), and DPI solutions that track ``pseudo-sessions'' have less material to work with, as each such pseudo-session has a lower bandwidth.

\item Finally, another future line of research is designing a new handshake protocol which is both secure against active probing while better utilizing the properties of DNS communication.

\end{itemize}

\bibliographystyle{plain}
\bibliography{Article}

\begin{thebibliography}{10}

\bibitem{ActiveProbing}
{Active Probing}.
\newblock \url{https://www.cs.princeton.edu/~rensafi/projects/active-probing/}.
\newblock [Online; accessed 1-Aug-2017].

\bibitem{AlexaTop}
{Alexa: The top 500 sites on the web}.
\newblock \url{https://www.alexa.com/topsites}.
\newblock [Online; accessed 23-Nov-2017].

\bibitem{Base58}
{Base58}.
\newblock \url{https://en.wikipedia.org/wiki/Base58}.
\newblock [Online; accessed 10-Aug-2017].

\bibitem{DomainFronting}
{Blocking-resistant communication through domain fronting}.
\newblock \url{https://www.bamsoftware.com/papers/fronting/}.
\newblock [Online; accessed 15-Nov-2017].

\bibitem{VPNBlockChina}
{China Tells Carriers to Block Access to Personal VPNs by February }.
\newblock
  \url{https://www.bloomberg.com/news/articles/2017-07-10/china-is-said-to-order-carriers-to-bar-personal-vpns-by-february}.
\newblock [Online; accessed 16-Aug-2017].

\bibitem{DNSMorphSourceCode}
{DNS-Morph source code, Github}.
\newblock Note: Please contact the ``Corresponding Author'' to get access to
  the Github repository.

\bibitem{GuyDNS}
{DNS part 2: visualization}.
\newblock \url{http://armatum.com/blog/2009/dns-part-ii/}.
\newblock [Online; accessed 21-Aug-2017].

\bibitem{Dns2tcp}
{Dns2tcp}.
\newblock \url{https://www.aldeid.com/wiki/Dns2tcp}.
\newblock [Online; accessed 20-Aug-2017].

\bibitem{Dnscat}
{Dnscat}.
\newblock \url{https://wiki.skullsecurity.org/Dnscat}.
\newblock [Online; accessed 20-Aug-2017].

\bibitem{dnslib}
{dnslib: A library to encode/decode DNS wire-format packets}.
\newblock \url{https://pypi.python.org/pypi/dnslib}.
\newblock [Online; accessed 6-Oct-2017].

\bibitem{FirstTor}
{First Tor Release}.
\newblock \url{http://archives.seul.org/or/dev/Sep-2002/msg00019.html}.
\newblock [Online; accessed 19-Aug-2016].

\bibitem{FteEvaluation}
{FTE Transport Evaluation}.
\newblock
  \url{https://trac.torproject.org/projects/tor/wiki/doc/PluggableTransports/FteEvaluation}.
\newblock [Online; accessed 25-Sep-2016].

\bibitem{PluggableTransportsList}
{Full List of Pluggable Transports}.
\newblock
  \url{https://trac.torproject.org/projects/tor/wiki/doc/PluggableTransports/list}.
\newblock [Online; accessed 20-Aug-2017].

\bibitem{NtorhHandshake}
{Improved circuit-creation key exchange}.
\newblock
  \url{https://gitweb.torproject.org/torspec.git/tree/proposals/216-ntor-handshake.txt}.
\newblock [Online; accessed 10-Sep-2016].

\bibitem{DNSExfiltration}
{INTRODUCTION TO DNS DATA EXFILTRATION }.
\newblock
  \url{https://blogs.akamai.com/2017/09/introduction-to-dns-data-exfiltration.html}.
\newblock [Online; accessed 21-Aug-2018].

\bibitem{iodine}
{iodine}.
\newblock \url{http://code.kryo.se/iodine/}.
\newblock [Online; accessed 20-Aug-2017].

\bibitem{KnockingBrigesDoors}
{Knock Knock Knockin' on Bridges' Doors}.
\newblock
  \url{https://blog.torproject.org/blog/knock-knock-knockin-bridges-doors}.
\newblock [Online; accessed 7-Sep-2016].

\bibitem{Libprotoident}
{Libprotoident}.
\newblock \url{https://research.wand.net.nz/software/libprotoident.php}.
\newblock [Online; accessed 25-Aug-2018].

\bibitem{Meek}
{Meek}.
\newblock \url{https://trac.torproject.org/projects/tor/wiki/doc/meek}.
\newblock [Online; accessed 22-Aug-2016].

\bibitem{MeekEvaluation}
{Meek Transport Evaluation}.
\newblock
  \url{https://trac.torproject.org/projects/tor/wiki/doc/PluggableTransports/MeekEvaluation}.
\newblock [Online; accessed 25-Sep-2016].

\bibitem{DNSLength2}
{More fun with DNS packet captures}.
\newblock
  \url{https://www.coverfire.com/archives/2008/07/28/more-fun-with-dns-packet-captures/}.
\newblock [Online; accessed 3-Jun-2017].

\bibitem{DNSMalwareCandC}
{Morto worm sets a (DNS) record}.
\newblock
  \url{https://www.symantec.com/connect/blogs/morto-worm-sets-dns-record}.
\newblock [Online; accessed 20-Aug-2017].

\bibitem{DNSMalwareExfiltration}
{Multigrain - Point of Sale Attackers Make an Unhealthy Addition to the
  Pantry}.
\newblock
  \url{https://www.fireeye.com/blog/threat-research/2016/04/multigrain_pointo.html}.
\newblock [Online; accessed 20-Aug-2017].

\bibitem{Namecheap}
{Namecheap: domain name registrar and web hosting company}.
\newblock \url{https://www.namecheap.com/}.
\newblock [Online; accessed 20-Nov-2017].

\bibitem{nDPI}
{nDPI - Open and Extensible LGPLv3 Deep Packet Inspection Library}.
\newblock \url{https://www.ntop.org/products/deep-packet-inspection/ndpi/}.
\newblock [Online; accessed 25-Nov-2017].

\bibitem{Netspark}
{Netspark}.
\newblock \url{http://netspark.com/}.
\newblock [Online; accessed 25-Nov-2017].

\bibitem{obfs3-spec}
{Obfs3 Protocol Specification}.
\newblock
  \url{https://github.com/NullHypothesis/obfsproxy/blob/master/doc/obfs3/obfs3-protocol-spec.txt}.
\newblock [Online; accessed 20-Sep-2016].

\bibitem{obfs4-spec}
{Obfs4 Protocol Specification}.
\newblock
  \url{https://github.com/Yawning/obfs4/blob/master/doc/obfs4-spec.txt}.
\newblock [Online; accessed 22-Aug-2016].

\bibitem{ObfsProxySourceCodeGo}
{ObfsProxy Source Code - Go}.
\newblock \url{https://gitweb.torproject.org/pluggable-transports/obfs4.git}.
\newblock [Online; accessed 25-Aug-2016].

\bibitem{ObfsProxySourceCodePython}
{ObfsProxy Source Code - Python}.
\newblock
  \url{https://gitweb.torproject.org/pluggable-transports/obfsproxy.git}.
\newblock [Online; accessed 25-Aug-2016].

\bibitem{DNSDomainNames}
{RFC 1123: Requirements for Internet Hosts -- Application and Support}.
\newblock \url{https://tools.ietf.org/html/rfc1123}.
\newblock [Online; accessed 3-Jun-2017].

\bibitem{SOCKS}
{RFC 1928: SOCKS Protocol}.
\newblock \url{https://www.ietf.org/rfc/rfc1928.txt}.
\newblock [Online; accessed 22-Aug-2016].

\bibitem{HMACRFC}
{RFC 2104: HMAC - Keyed-Hashing for Message Authentication}.
\newblock \url{https://www.ietf.org/rfc/rfc2104.txt}.
\newblock [Online; accessed 20-Nov-2017].

\bibitem{Base32RFC}
{RFC 3548: The Base16, Base32, and Base64 Data Encodings}.
\newblock \url{https://tools.ietf.org/html/rfc3548.html}.
\newblock [Online; accessed 6-Feb-2017].

\bibitem{RetransmissionTimer}
{RFC 6298: Computing TCP's Retransmission Timer}.
\newblock \url{https://tools.ietf.org/html/rfc6298}.
\newblock [Online; accessed 10-Jun-2017].

\bibitem{UDPHeader}
{RFC 768: User Datagram Protocol}.
\newblock \url{https://www.ietf.org/rfc/rfc768.txt}.
\newblock [Online; accessed 3-Jun-2017].

\bibitem{InternetProtocol}
{RFC 791: Internet Protocol - Protocol Specification}.
\newblock \url{https://tools.ietf.org/html/rfc791#page-11}.
\newblock [Online; accessed 3-Jun-2017].

\bibitem{Skype}
{Skype: online text message and video chat services}.
\newblock \url{https://www.skype.com}.
\newblock [Online; accessed 25-Jul-2017].

\bibitem{Snowflake}
{Snowflake}.
\newblock \url{https://trac.torproject.org/projects/tor/wiki/doc/Snowflake}.
\newblock [Online; accessed 22-Aug-2018].

\bibitem{UniformDH}
{The UniformDH scheme - Ian Goldberg}.
\newblock
  \url{https://lists.torproject.org/pipermail/tor-dev/2012-December/004245.html}.
\newblock [Online; accessed 20-Sep-2016].

\bibitem{BridgeDB}
{Tor bridges database}.
\newblock \url{https://bridges.torproject.org/}.
\newblock [Online; accessed 30-Oct-2017].

\bibitem{TorClientSoftware}
{Tor Client Software}.
\newblock \url{https://www.torproject.org/download/download}.
\newblock [Online; accessed 10-Sep-2016].

\bibitem{TorMetrics}
{Tor Metrics}.
\newblock \url{https://metrics.torproject.org/}.
\newblock [Online; accessed 16-Aug-2018].

\bibitem{PluggableTransportsSpec}
{Tor: Pluggable Transports Specification}.
\newblock \url{https://gitweb.torproject.org/torspec.git/tree/pt-spec.txt}.
\newblock [Online; accessed 22-Aug-2016].

\bibitem{TorWebsite}
{Tor Website}.
\newblock \url{https://www.torproject.org/}.
\newblock [Online; accessed 25-Aug-2016].

\bibitem{WhatsApp}
{WhatsApp Messenger}.
\newblock \url{https://www.whatsapp.com/}.
\newblock [Online; accessed 25-Jul-2017].

\bibitem{GFCDNS}
Anonymous.
\newblock Towards a comprehensive picture of the great
  firewall{\textquoteright}s {DNS} censorship.
\newblock In {\em 4th {USENIX} Workshop on Free and Open Communications on the
  Internet ({FOCI} 14)}, pages 445--458, San Diego, CA, 2014. {USENIX}
  Association.

\bibitem{ServicesPorts}
Internet Assigned~Numbers Authority.
\newblock {Service Name and Transport Protocol Port Number Registry}.
\newblock
  \url{https://www.iana.org/assignments/service-names-port-numbers/service-names-port-numbers.xhtml}.
\newblock [Online; accessed 14-Jan-2017].

\bibitem{Elligator}
Daniel~J. Bernstein, Mike Hamburg, Anna Krasnova, and Tanja Lange.
\newblock Elligator: Elliptic-curve points indistinguishable from uniform
  random strings.
\newblock Cryptology ePrint Archive, Report 2013/325, 2013.
\newblock \url{http://eprint.iacr.org/2013/325}.

\bibitem{InternetworkingBook}
Douglas~E. Comer.
\newblock {\em Internetworking with TCP/IP, Volume 1: Principles, Protocols,
  and Architectures, Fourth Edition}.
\newblock Prentice Hall PTR, Upper Saddle River, NJ, USA, 4th edition, 2000.

\bibitem{FTE}
Kevin~P. Dyer, Scott~E. Coull, Thomas Ristenpart, and Thomas Shrimpton.
\newblock Protocol misidentification made easy with format-transforming
  encryption.
\newblock In {\em Proceedings of the 2013 ACM SIGSAC Conference on Computer 38;
  Communications Security}, CCS '13, pages 61--72, New York, NY, USA, 2013.
  ACM.

\bibitem{Marionette}
Kevin~P. Dyer, Scott~E. Coull, and Thomas Shrimpton.
\newblock Marionette: A programmable network traffic obfuscation system.
\newblock In {\em 24th {USENIX} Security Symposium ({USENIX} Security 15)},
  pages 367--382, Washington, D.C., 2015. {USENIX} Association.

\bibitem{GFC2015}
Roya Ensafi, David Fifield, Philipp Winter, Nick Feamster, Nicholas Weaver, and
  Vern Paxson.
\newblock Examining how the great firewall discovers hidden circumvention
  servers.
\newblock In {\em Proceedings of the 2015 ACM Conference on Internet
  Measurement Conference}, IMC '15, pages 445--458, New York, NY, USA, 2015.
  ACM.

\bibitem{fry2009security}
C.~Fry and M.~Nystrom.
\newblock {\em Security Monitoring: Proven Methods for Incident Detection on
  Enterprise Networks}.
\newblock O'Reilly Media, 2009.

\bibitem{Parrot}
Amir Houmansadr, Chad Brubaker, and Vitaly Shmatikov.
\newblock The parrot is dead: Observing unobservable network communications.
\newblock In {\em 2013 {IEEE} Symposium on Security and Privacy, {SP} 2013,
  Berkeley, CA, USA, May 19-22, 2013}, pages 65--79, 2013.

\bibitem{DNSLength1}
A.A. Manaf, A.~Zeki, M.~Zamani, S.~Chuprat, and E.~El-Qawasmeh.
\newblock {\em Informatics Engineering and Information Science}.
\newblock Springer Berlin Heidelberg, 2011.

\bibitem{Paxson}
Vern Paxson, Mihai Christodorescu, Mobin Javed, Josyula Rao, Reiner Sailer,
  Douglas~Lee Schales, Mark Stoecklin, Kurt Thomas, Wietse Venema, and Nicholas
  Weaver.
\newblock Practical comprehensive bounds on surreptitious communication over
  {DNS}.
\newblock In {\em Presented as part of the 22nd {USENIX} Security Symposium
  ({USENIX} Security 13)}, pages 17--32, Washington, D.C., 2013. {USENIX}.

\bibitem{GFC2012}
Philipp Winter and Jedidiah R~. Crandall.
\newblock The {G}reat {F}irewall of {C}hina: How it blocks {T}or and why it is
  hard to pinpoint.
\newblock {\em USENIX ;login:}, 37(6):42--50, 2012.

\bibitem{ScrambleSuit}
Philipp Winter, Tobias Pulls, and J{\"{u}}rgen Fu{\ss}.
\newblock Scramblesuit: a polymorphic network protocol to circumvent
  censorship.
\newblock In {\em Proceedings of the 12th annual {ACM} Workshop on Privacy in
  the Electronic Society, {WPES} 2013, Berlin, Germany, November 4, 2013},
  pages 213--224, 2013.

\end{thebibliography}

\end{document}